\documentclass[aps,twocolumn,floats,nofootinbib,pre]{revtex4}
\usepackage{graphics,graphicx,epsfig}
\usepackage{amssymb,color}
\usepackage{epsf,epstopdf,wrapfig}
\usepackage {amsmath}

\newcommand{\beq}{\begin{equation}}
\newcommand{\eeq}{\end{equation}}
\newcommand{\beqn}{\begin{eqnarray}}
\newcommand{\eeqn}{\end{eqnarray}}

\begin{document}

\title{PCA meets RG}

\author{Serena Bradde$^a$ and William Bialek$^{a,b}$}

\affiliation{$^a$Initiative for the Theoretical Sciences, The Graduate Center, City University of New York, 365 Fifth Ave., New York, New York 10016\\
$^b$Joseph Henry Laboratories of Physics, and Lewis--Sigler Institute for Integrative Genomics, Princeton University, Princeton NJ 08544}

\date{\today}

\begin{abstract}
A system with many degrees of freedom can be characterized by a covariance matrix;  principal components analysis (PCA) focuses on the eigenvalues of this matrix, hoping to find a lower dimensional description.  But when the spectrum is nearly continuous, any distinction between components that we keep and those that we ignore becomes arbitrary; it then is natural to ask what happens as we vary this arbitrary cutoff.  We argue that this problem is analogous to the momentum shell renormalization group (RG).  Following this analogy, we can define relevant and irrelevant operators, where the role of dimensionality is played by properties of the eigenvalue density.  These results also suggest an approach to the analysis of real data.  As an example, we study neural activity in the vertebrate retina as it responds to naturalistic movies, and find evidence of behavior  controlled by a nontrivial fixed point.  Applied to financial data, our analysis separates modes dominated by sampling noise from a smaller but still macroscopic number of modes described by a non--Gaussian distribution.
\end{abstract}

\maketitle

\section{Introduction}

Many of the most interesting phenomena in the world around us emerge from interactions among many degrees of freedom.  In the era of ``big data,'' we are encouraged to think about this more explicitly, describing the state of the system as a point in a space with many dimensions:  the state of a cell is defined by the expression level of many genes, the state of a financial market is defined by the prices of many stocks, and so on.  One  approach to the analysis of these high dimensional data is to look for a linear projection onto a lower dimensional space.  Quantitatively, the best projection is found by diagonalizing the covariance matrix, which decomposes the variations into modes that are independent at second order; these are the principal components (PCs), and the method is called principal components analysis (PCA).\footnote{The idea of PCA goes back at least to the start of the twentieth century \cite{pca1}. For a brief modern summary, see Ref \cite{pca2}.} In favorable cases, very few modes will capture most of the variance, but it is much more common to find that the eigenvalues of the covariance matrix form a more continuous spectrum, so that any sharp  division between important and unimportant dimensions would be arbitrary.

For physical systems in thermal equilibrium, it again is the case that the most interesting phenomena emerge from interactions among many degrees of freedom,  but here we have a quantitative language for describing this emergence.  In the classical view, we make precise models of the interactions on a microscopic scale, and then statistical mechanics is about calculating the implications of these interactions for the macroscopic behavior of matter.   In the modern view, we admit that our microscopic description itself is approximate, incorporating ``effective interactions'' mediated by degrees of freedom that we might not want to describe explicitly, and that the distance scale at which we draw the boundary between explicit and implicit description also is arbitrary.  Attention shifts from the precise form of our model to the way in which this model evolves as we move the boundary between degrees of freedom that we describe and those that we ignore \cite{wilson_79}; the evolution through the space of possible models is described by the renormalization group (RG).  A central result of the RG is that many detailed features of models on a microscopic scale disappear as we coarse--grain our description out to the macroscopic scale, and that in many cases we are left with only a few terms in our models, the ``relevant operators.''  Thus, some of the success of simple models in describing the world comes not from an inherent simplicity, but rather from the fact that macroscopic behaviors are insensitive to most microscopic details (irrelevant operators).  

The RG approach to statistical physics suggests that systems in which PCA fails to yield a clean separation between high variance and low variance modes may nonetheless be simplified.  Indeed, in a system where the many degrees of freedom live on a lattice, with translation invariant interactions, the  principal components are Fourier modes, and typically we find that the variance of each mode decreases monotonically but smoothly with wavelength.  In the momentum space implementation of RG \cite{wilson+kogut_74}, we put a cutoff on the wavelength, and ask what happens to the joint distribution of the remaining variables as we move this cutoff, averaging over the short wavelength  modes.  In this language, the RG is about what happens as we vary the arbitrary distinction between high variance PCs that we keep, and low variance PCs that we ignore.    The goal of this paper is to clarify this connection between PCA and RG, so that we can construct RG approaches to  more complex, high dimensional systems.  

\section{Historical note}

This paper has been written for a volume dedicated to the memory of Leo Kadanoff.  As has been described many times, the modern development of the RG began with Leo's intuitive construction of ``block spins,'' in which he made explicit the idea of averaging over the fluctuations that occur on short wavelengths \cite{kadanoff_66}.    Later in his life, Kadanoff worked on more complex problems, from the dynamics of cities \cite{kadanoff_71,kadanoff+weinblatt_72} to patterns \cite{bensimon+al_86}, chaos \cite{halsey+al_86}, and singularities \cite{constantin+kadanoff_90,bertozzi+al_94} in fluid flows, and more \cite{coppersmith+al_99,povinelli+al_99}.  Beyond his own work, he was a persistent advocate for the physics community's exploration of complex systems, including biological systems.  We have benefited, directly and indirectly, from his enthusiasm, as well as being inspired by his example.

Among Leo's last papers are a series of historical pieces reflecting on his  role in the development of the RG, and on  statistical physics more generally \cite{kadanoff_09,kadanoff_13,kadanoff_14,kadanoff_15}.  Although much can be said about these papers, surely one message is that the physicist's persistent search for simplification has been rewarded, time and again.    We offer this paper in that spirit, as we try to carry Kadanoff's intuition about thinning out microscopic degrees of freedom away from its origins in systems with local interactions.

\section{Correlation spectra and effective dimensions}

Let us imagine that the system we are studying is described by a set of variables $\phi_1, \, \phi_2, \, \cdots ,\, \phi_N \equiv \{\phi_{\rm i}\}$, where the dimensionality $N$ is large.  For the purposes of this discussion, ``describing the system'' means  writing down the joint probability of all $N$ variables, $P(\{\phi_{\rm i}\})$.  For simplicity we define these variables so that they have zero mean, and we'll assume that positive and negative fluctuations are equally likely (though this is not essential).    We start with the guess that the fluctuations are nearly Gaussian, so we can write
\begin{equation}
P(\{\phi_{\rm i}\}) = \frac{1}{Z}\exp\left[ - \frac{1}{2}\sum_{{\rm i},{\rm j}} \phi_{\rm i} K_{\rm ij} \phi_{\rm j} - \frac{1}{4!} g \sum_{\rm i}  \phi_{\rm i}^4 + \cdots \right] ,
\label{P1}
\end{equation}
where the coefficient $g$ allow us to describe weak kurtosis of the random variables.  It may be useful to note that the probability distribution in Eq (\ref{P1}) is the maximum entropy, and hence least structured,  model consistent with the full covariance matrix and the mean kurtosis  of all the variables; in this sense it is a minimal model.  Much of what we will say here can generalized to the case where each variable has a different kurtosis, so there is a distinct $g_{\rm i}$ associated with each term $\phi_{\rm i}^4$.

If $g=0$, we are describing a system in which fluctuations are Gaussian, and in this limit the matrix $K_{\rm ij}$ is the inverse of the covariance matrix
\begin{equation}
C_{\rm ij} = \langle \phi_{\rm i}\phi_{\rm j}\rangle .
\end{equation}
In the conventional application of renormalization group ideas, we can classify non--Gaussian terms as relevant or irrelevant: as we coarse--grain our description from microscopic to macroscopic scales, do departures from a Gaussian distribution become more or less important? Our first goal is to show how we can export this idea to the more general setting, where the kernel $K_{\rm ij}$ does not have any symmetries such as translation invariance.  To do this, we start near $g=0$, and work in perturbation theory.

It is useful to write the eigenvalues $\lambda_\mu$ and eigenvectors $\{u_{\rm i}(\mu )\}$ of the matrix $K$,
\begin{equation}
\sum_{\rm j} K_{\rm ij} u_{\rm j}(\mu )  = \lambda_\mu u_{\rm i}(\mu ),
\end{equation}
so that the variations in $\{\phi_{\rm i}\}$ can be decomposed into modes $\{\tilde\phi_\mu\}$,
\begin{equation}
\phi_{\rm i} = \sum_\mu u_{\rm i} (\mu )\tilde\phi_\mu ;
\end{equation}
 if $g = 0$ then these modes are exactly the principal components.
The Gaussian term becomes
\begin{equation}
  \frac{1}{2}\sum_{{\rm i},{\rm j}} \phi_{\rm i} K_{\rm ij} \phi_{\rm j}  =   \frac{1}{2}\sum_\mu \lambda_\mu \tilde\phi_\mu^2 ,
  \end{equation}
and hence at $g_{\rm i}\rightarrow 0$ the variance of each mode is given by $\langle \tilde\phi_\mu^2\rangle = 1/\lambda_\mu$.  The average variance of the individual variables is
\begin{equation}
\frac{1}{N}\sum_{\rm i}\langle \phi_{\rm i}^2\rangle = \frac{1}{N}\sum_\mu \frac{1}{\lambda_\mu} \rightarrow \int_0^\Lambda d\lambda\, \rho(\lambda ) \frac{1}{\lambda},
\label{meanvar1}
\end{equation}
where in the last step we introduce the distribution 
\begin{equation}
\rho (\lambda ) = \frac{1}{N}\sum_\mu \delta (\lambda - \lambda_\mu ),
\end{equation}
which becomes smooth in the limit of large $N$, and we note explicitly that there is a largest eigenvalue $\Lambda$.

The essential idea is  to eliminate the modes that have small variance.  This corresponds to restricting our attention only to modes with $\lambda$ {\em less} than some cutoff.  Equivalently, it corresponds to decreasing the limit $\Lambda$ on the integral over eigenvalues, e.g. in Eq (\ref{meanvar1}).  This reduces the total variance,  but it is natural to choose units in which the variance is fixed, and  this implies that as we change the cutoff $\Lambda$ we have to rescale the values of $\phi_{\rm i}$.  So we  replace $\phi_{\rm i} \rightarrow z_\Lambda \phi_{\rm i}$, and we can determine this scale factor by insisting that the mean variance stay fixed.  Again, we are working at small $g$, so we do this calculation at $g=0$:
\begin{eqnarray}
0 &=& \frac{d}{d\Lambda} \left[ \frac{1}{N}\sum_{\rm i}\langle (z_\Lambda\phi_{\rm i})^2\rangle\right]\\
&=& \frac{d}{d\Lambda} \left[z_\Lambda^2 \int_0^\Lambda d\lambda\,  \frac{\rho(\lambda )}{\lambda}\right]\\
\Rightarrow \frac{d\ln z_\Lambda}{d\ln \Lambda} &=& - \frac{1}{2} \rho(\Lambda ) \left[ \int_0^\Lambda d\lambda\, \rho(\lambda ) \frac{1}{\lambda}\right]^{-1} .
\end{eqnarray}

When we reduce the cutoff, we also reduce the number of degrees of freedom in the system.  The average of the quadratic term in the (log) probability distribution is automatically proportional to this effective number of degrees of freedom, 
\begin{equation}
N_{\rm eff} = N\int_0^\Lambda d\lambda \,\rho(\lambda ),
\end{equation}
and this insures, for example, that the entropy of the probability distribution will be proportional to $N_{\rm eff}$ (extensivity). 
To be sure that this works also for the quartic terms, we write
\begin{equation}
N  g  \left[{1\over N}\sum_{\rm i} \phi_{\rm i}^4\right] = N_{\rm eff} \tilde g  \left[{1\over N}\sum_{\rm i} \left(z_\Lambda \phi_{\rm i}\right)^4\right],
\end{equation}
which defines the effective coupling constant
\begin{equation}
 \tilde g = z_{\Lambda}^{-4} g \frac{N}{N_{\rm eff}} .
\end{equation}
Now the scaling of the coefficient $\tilde g$  is given by
\begin{equation}
\frac{d\ln \tilde g}{d\ln\Lambda} = \rho(\Lambda) \left[  \frac{2}{\int_0^\Lambda d\lambda\, \rho(\lambda ) \frac{1}{\lambda}} - \frac{\Lambda}{\int_0^\Lambda d\lambda \,\rho(\lambda )} \right] .
\label{dG4}
\end{equation}
Since this is the difference between two positive terms, we can find either sign for the result.

If the scaling function ${d\ln \tilde g}/{d\ln\Lambda}$ is positive, then as we decrease the cutoff $\Lambda$  and thus average over more and more of the low variance modes, any small quartic term $\tilde g$ will become still smaller, and hence the distribution approaches a Gaussian.  This seems to make sense, since when we project onto a (much) lower dimensional space, each of the variables that remains is a weighted sum of many of the original variables, and we might expect the central limit theorem to enforce approximate Gaussianity of the resulting distribution.    But if the scaling function ${d\ln \tilde g }/{d\ln\Lambda} < 0$, then as we average over more and more of the lower variance modes, the quartic term becomes more and more important to the structure of the distribution.    To use the language of the RG, under these conditions the quartic term is a relevant operator.  

If we consider the case where the density of eigenvalues is a power law, $\rho = B\lambda^{\alpha -1}$, we find
\begin{equation}
\frac{d\ln \tilde g }{d\ln\Lambda} = \alpha - 2.
\label{dG4b}
\end{equation}
Thus, the spectral density of eigenvalues determines the relevance of non--Gaussian terms in the distribution.

In the conventional field theoretic examples, where the variables $\phi$ live at positions $\mathbf x$ in a $D$ dimensional Euclidean space, the correlations come from a  ``kinetic energy'' term that enforces similarity among neighbors, 
\begin{equation}
 \frac{1}{2}\sum_{{\rm i},{\rm j}} \phi_{\rm i} K_{\rm ij} \phi_{\rm j}  \rightarrow \frac{1}{2}\int d^D x \left[ {\mathbf\nabla} \phi({\mathbf x}) \right]^2 .
\end{equation}
The eigenvectors of $K$ are  Fourier modes,  indexed by a wave vector $\mathbf k$, with eigenvalues $\lambda = |\mathbf k|^2$.  If the original variables are on a lattice with linear spacing $a$,  then there is a maximum eigenvalue $\Lambda \sim (\pi/a)^2$, and  the density 
\begin{equation}
\rho(\lambda ) \propto \int d^D k \,\delta\left( \lambda - |\mathbf k|^2 \right) \propto  \lambda^{{D/2}-1} ,
\label{rhoD}
\end{equation}
corresponding to $\alpha = D/2$.  From Eq (\ref{dG4b}) we find 
\begin{equation}
\frac{d\ln \tilde g }{d\ln\Lambda} =  \frac{D}{2} - 2 = \frac{1}{2}\left( D - 4\right).
\end{equation}
The quartic term is relevant if ${d\ln \tilde g }/{d\ln\Lambda} < 0$, which corresponds to $D < 4$, as is well known from the conventional RG analysis \cite{wilson+fisher_72,amit+martin-mayor}; the extra factor of $1/2$ arises because $\Lambda$ is a cutoff on the eigenvalue, which is the square of the wavevector.  

Everything that we have said here can be carried over to the case where each variable is associated with a different coupling $g_{\rm i}$ in our original model, Eq (\ref{P1}).  This corresponds to a maximum entropy description that captures the pairwise correlations among the variables and the kurtosis of each individual variable.  The relevance or irrelevance of each term is controlled, in the same way, by the eigenvalue spectrum.  If we include terms $\sim \phi_{\rm i}^n$, allowing us to match higher moments of the marginal distributions for each $\phi_{\rm i}$, then as usual these terms are less relevant at larger $n$.

\section{Can we find fixed points?}

Thus far our analysis has been confined to ``power counting.''    The next step is to   integrate out the the low variance degrees of freedom and compute corrections to the coupling constants that are beyond those generated from the spectrum of eigenvalues itself.  Since we can think about discrete modes, we can write $\phi_{\rm i} \rightarrow \phi_{\rm i} + u_{\rm i}  \psi$, where $\psi$ is the variable describing fluctuations in the ``last mode'' that we have kept in our description, and we want to average over these fluctuations.  In the limit of small $g$,  $\psi$ is Gaussian with $\langle \psi^2\rangle = 1/\Lambda$, and we find
\begin{widetext}
\begin{eqnarray}\label{Hint}
\exp\left[ - \frac{\tilde g}{4!}  \frac{ N_{\rm eff}}{N}\sum_{\rm i} (z_\Lambda \phi_{\rm i})^4\right]
&&\rightarrow 
{\bigg\langle}\exp\left[ -  \frac{\tilde g}{4!} \frac{ N_{\rm eff}}{N} \sum_{\rm i}  z_\Lambda^4 (\phi_{\rm i} + u_{\rm i}  \psi)^4\right]{\bigg\rangle}\\
&&\hskip -1.25in =  
\exp\left[ - \frac{\tilde g}{2}   \frac{ N_{\rm eff}}{N} \sum_{\rm i}  z_\Lambda^4 {{u_{\rm i}^2 }\over \Lambda} \phi_{\rm i}^2
 - \frac{\tilde g}{4!} \frac{ N_{\rm eff}}{N} \sum_{\rm i} (z_\Lambda \phi_{\rm i})^4\
 + \frac{1}{2}\left( \frac{\tilde g N_{\rm eff}}{4! N}\right)^2   \sum_{{\rm i},{\rm j}} \frac{ z_\Lambda^8}{\Lambda^2}  \left( 72 \phi_{\rm i}^2 \phi_{\rm j}^2 u_{\rm i}^2 u_{\rm j}^2 + 96 \phi_{\rm i}^3 \phi_{\rm j} u_{\rm i} u_{\rm j}^3\right) + \cdots \right] . \nonumber\\
 &&
\end{eqnarray}
\end{widetext}

The first term is a correction to the matrix $K$, analogous to a mass renormalization. In the general case we not only get corrections to the coefficient of $\phi_{\rm i}^4$, we also generate terms $\sim \phi_{\rm i}^2 \phi_{\rm j}^2$ and $\sim \phi_{\rm i}^3 \phi_{\rm j}$.
As in the standard discussion, we will assume that the fields $\phi_{\rm i}$ are ``slowly varying'' functions of their index.  More precisely, we will expand the correction terms around the point where $\phi_{\rm i} = \phi_{\rm j}$, and for now we drop the gradient--like terms $\sim (\phi_{\rm i} - \phi_{\rm j})$, $\sim (\phi_{\rm i} - \phi_{\rm j})^2$, \dots.
In this approximation, we have 
\begin{eqnarray}
\sum_{{\rm i},{\rm j}}  \phi_{\rm i}^2 \phi_{\rm j}^2 u_{\rm i}^2 u_{\rm j}^2  &\approx&
 \sum_{\rm i} \phi_{\rm i}^4 u_{\rm i}^2 \sum_{\rm j} u_{\rm j}^2\\
&=&  \sum_{\rm i} \phi_{\rm i}^4 u_{\rm i}^2 \approx {1\over N} \sum_{\rm i} \phi_{\rm i}^4 ,
\end{eqnarray}
where in the last step we again use the slow variation of $\phi_{\rm i}$ to replace $u_{\rm i}^2$ with its average.
In the same approximation, the term $\sim u_{\rm i} u_{\rm j}^3$ vanishes.  The net result is that
\begin{equation}
\tilde g \rightarrow \tilde g - \frac{3}{2} {\tilde g}^2 \frac{N_{\rm eff}}{N} \frac{z_\Lambda^4}{N} \frac{1}{\Lambda^2} .
\label{dgA}
\end{equation}
This is the change in coupling associated with integrating out one mode, which corresponds to a change in the cutoff such that
$-d\Lambda \rho(\Lambda) N = 1$, so we can rewrite Eq (\ref{dgA}) as
\begin{equation}
\frac {d \ln \tilde g}{d\ln \Lambda} =  \frac{3}{2} {\tilde g} \frac{N_{\rm eff}}{N}  z_\Lambda^4 \frac{ \rho(\Lambda)}{ \Lambda} .
\label{dgB}
\end{equation}
Combining with the scaling behavior in Eq (\ref{dG4}), we find
\begin{equation}
\frac {d \ln \tilde g}{d\ln \Lambda} =  \rho(\Lambda)\left[\frac{2}{\int_0^\Lambda d\lambda\, \rho(\lambda ) \frac{1}{\lambda}} - \frac{\Lambda}{\int_0^\Lambda d\lambda \,\rho(\lambda )} +  \frac{3}{2} \frac{ {\tilde g} }{  \Lambda} \right].
\label{dgC}
\end{equation}
In the case where $\rho(\lambda ) \propto \lambda^{\alpha -1}$, this generates a fixed point $\tilde g_* \propto \alpha -2$, which is analogous to the Wilson--Fisher  fixed point $\tilde g_* \propto 4 - D$ \cite{wilson+fisher_72}.

The calculation we have done here is aimed at showing that the conventional analysis of fixed points can be carried over to this different setting, away from equilibrium statistical physics with local interactions.  We assume that this more complex setting allows for a richer variety of fixed points, which need to be explored.

\section{An approach to data analysis?}

These arguments suggest that, at least in perturbation theory, much of the apparatus of the renormalization group for translation invariant systems with local interactions can be carried over to  more complex systems.  We can define relevant and irrelevant operators, and there is a path to identifying fixed points.  The crucial role played by the dimensionality in systems with local interactions is played instead by the spectrum of the matrix $K$.  Perhaps most important is that we can carry over the {\em concept} of renormalization.

\begin{figure*}[t]
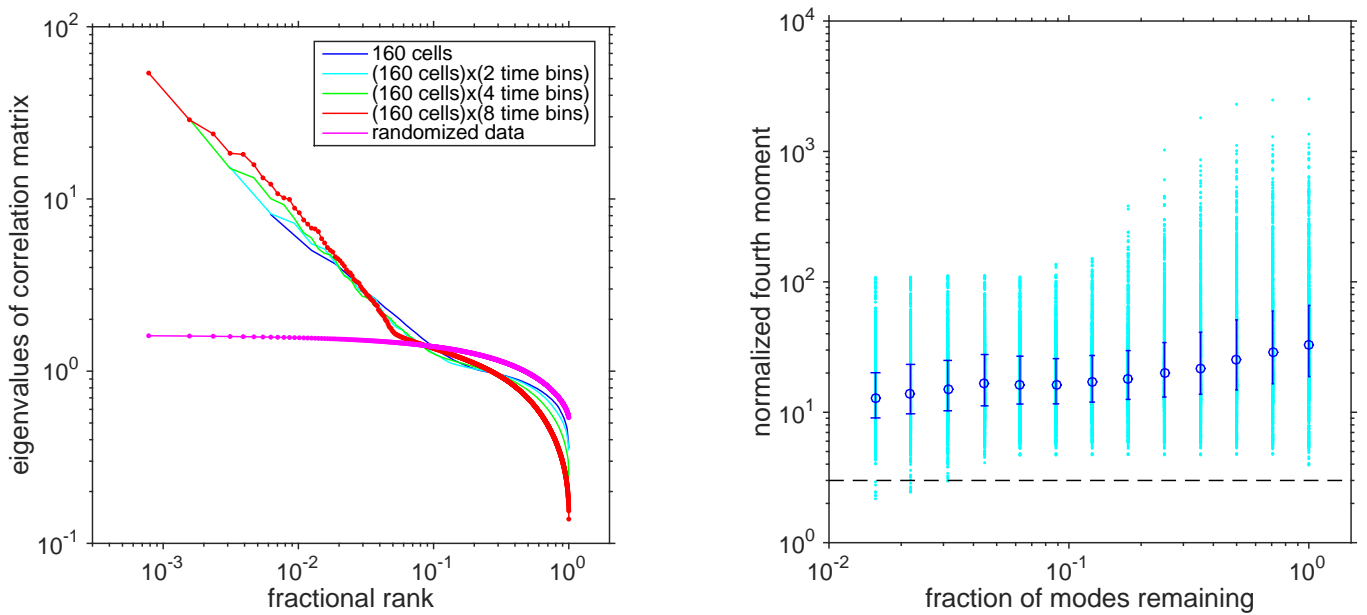

\includegraphics[width = 0.45\linewidth]{retina_spectrum}
\hfill
\includegraphics[width = 0.45\linewidth]{retina_M4}
\caption{Analysis of neural activity in the retina. States are defined by the patterns of spiking and silence in successive time bins from 160 neurons, as described in the text.  Left: Spectrum of eigenvalues of the correlation matrix, with states constructed from different numbers of time bins, plotted vs fractional mode number; results from randomized data shown for comparison.   Right:  Normalized fourth moments for each of the $8\times 160$ variables (cyan dots),   as a function of the fraction of modes remaining after coarse graining; blue circles show medians $\pm$ one quartile. \label{retinal_data}}
\end{figure*}

Faced with real data on a system with many degrees of freedom, we don't know the matrix $K$.  We do know that, if the system is close to being Gaussian, then $K$ is close to being the inverse of the covariance matrix $C$, which we can estimate from the data.  In systems with translation invariance, the eigenvectors of $C$ and $K$ are the same, which means that coarse graining by eliminating the modes with large eigenvalues of $K$ (momentum shells) is exactly the same as eliminating the modes with small eigenvalues of $C$.   Although this can't be true in general, we can try:

(1)  We examine the spectrum of the covariance matrix.  If a small number of eigenvalues are separated from the bulk, and capture most of the variance, then the system is genuinely low dimensional and we are done (PCA works).  More commonly, we find a continuum of eigenvalues, with no natural separation.

(2)  Power--law behavior in a rank--ordered plot of the eigenvalues is analogous to power--law correlation functions in the usual field theoretic or statistical physics examples.  In practice, however, it may be difficult to verify power--law behavior over a very wide range of scales.

(3)  We coarse grain our description by projecting out a fraction of the modes with the smallest eigenvalues of $C$.  In effect this replaces each variable $\phi_{\rm i}$ by an average over low variance details, in the spirit of the block spin construction.

(4) To follow the results of coarse graining, we can measure the moments of the local variables,  $\langle \phi_{\rm i}^n\rangle$, or even their full distribution, as was done long ago for Monte Carlo data by Binder \cite{binder_81}.

As a first  example, we have analyzed an experiment on the activity of 160 neurons in a small patch of the vertebrate retina as it responds to naturalistic movies \cite{tkacik+al_14}; a full description will be given elsewhere, but here we focus on our ability to detect a nontrivial renormalization group flow as we coarse--grain this system.  As in previous analyses of these data, we divide time into small bins (width $\Delta\tau = 20\,{\rm ms}$), and in each bin a single neuron either generates an action potential or remains silent, so that the natural local variables are binary before any coarse graining.  We can then take a state of the entire system to be the 160--dimensional vector of these binary variables, but we can also consider $T$ successive vectors in time, as in the ``time delayed embedding'' analysis of dynamical systems \cite{TD}:   increasing $T$ compensates for not observing directly all the relevant degrees of freedom in the system, and gives us access to a higher dimensional description.  Given the size of the data set, we can use $T=8$ without creating problems of undersampling, and this gives  us $N=1280$ dimensions.   Because the different neurons are different from one another, we normalize each variable $\phi_{\rm i}$ to have zero mean and unit variance, and so the covariance matrix is the matrix of correlation coefficients in the raw data.

As we can see at left in Fig \ref{retinal_data},  the eigenvalues of the correlation matrix have an essentially continuous spectrum, perhaps even showing hints of scale invariance.\footnote{In some contexts it would be more conventional to look at the distribution of eigenvalues, searching for modes with high variance that emerge clearly from a ``bulk'' that might be ascribed to sampling noise.    Plotting eigenvalues vs their rank, as we do here, provides a representation of the cumulative distribution of eigenvalues, and does not require us to make bins along the eigenvalue axis.   Rather than plotting from smallest to largest, we plot from largest to smallest, so that the spectra are more directly comparable to a plot of the susceptibility or propagator $G(k)$ vs momentum $k$ in the usual statistical physics examples.}   This spectral structure is well outside the range that would be generated by an equally large random sample from uncorrelated variables.

\begin{figure*}[t]
\includegraphics[width = 0.45\linewidth]{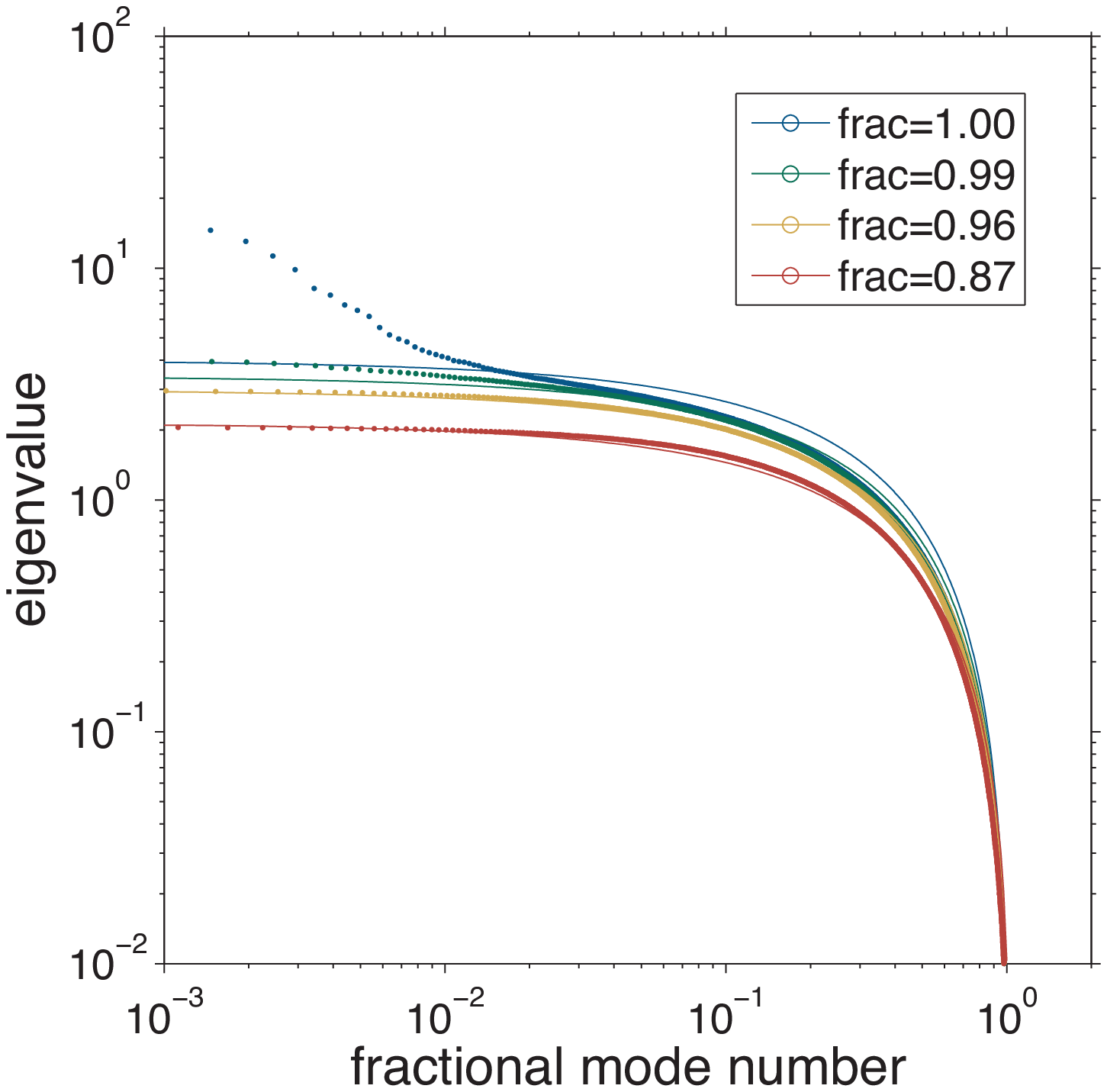}
\hfill
\includegraphics[width = 0.45\linewidth]{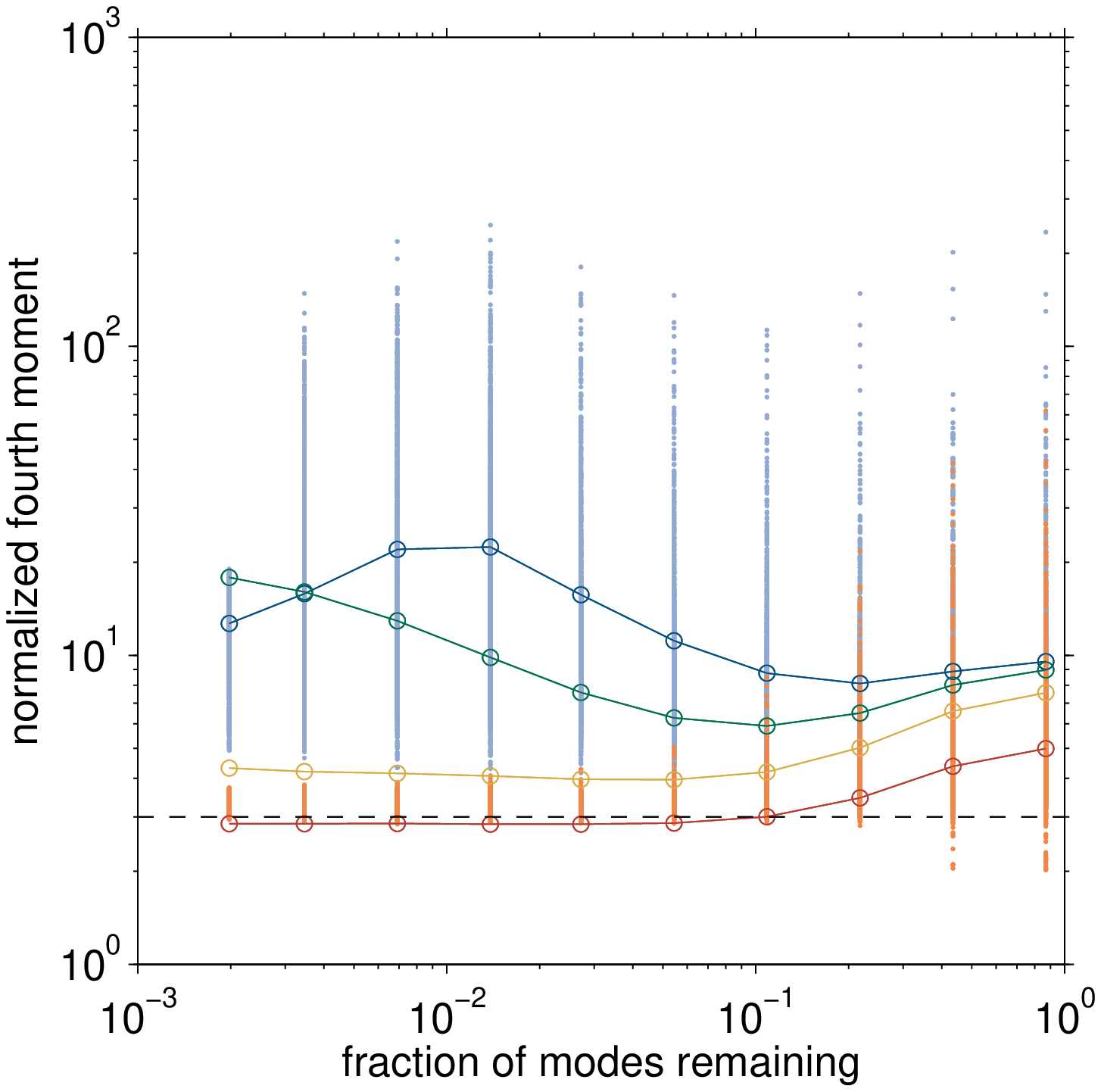}
\caption{Analysis of daily returns on f $N=2048$ assets in the NYSE for a period of $T=2356$ days. Left: Spectrum of eigenvalues of the correlation matrix; results are shown for all the data (blue), as well as cases in which we remove the the largest $1\%$ (green), $4\%$ (yellow), or $13\%$ (red) of the eigenvalues. Solid lines are theoretical expectations from the Marchenko--Pastur distribution  \cite{JPB}. Right: Normalized fourth moments for each of the $N=2048$ variables, as a function of the fraction of modes remaining after coarse graining. Blue and red dots are the normalized fourth moments  and the lines traces their mean. Colors code different cases where track all the modes, or first remove different fractions of the large variance modes, as in the analysis at left. \label{stock_data}}
\end{figure*}

Because the raw variables of this system are binary, the normalized fourth moments ($\langle \phi_{\rm i}^4\rangle/\langle\phi_{\rm i}^2\rangle^2$) can be large and vary substantially from neuron to neuron.  As we coarse--grain,  eliminating modes corresponding to small eigenvalues of the covariance matrix,  this variation is reduced, as shown at right in Fig \ref{retinal_data}.  More strikingly,  the normalized fourth moments are hardly varying as we move our cutoff beyond the first $\sim 90\%$ of the modes, and the median value is stabilizing  well above the value of 3 expected for a Gaussian distribution.    It is interesting that the range of scales (fraction of modes included) over which see an approximately fixed fourth moment is the same as the range over which see approximately power--law behavior of the eigenvalue spectrum.  These results suggest that the joint distribution of activity in this neural network is close to a nontrivial fixed point of the renormalization group transformation.  This is consistent with previous evidence that this system is close to a critical point in the thermodynamic sense \cite{mora+bialek_11,tkacik+al_15}, but the renormalization group analysis connects more fully to our understanding of criticality in equilibrium systems.

As a second example, we consider a set of 4000 assets traded on the New York Stock Exchange \cite{Marsili2002,Mantegna}.   In the data, spanning nearly ten years from 1 Jan 1990 through 30 Apr 1999,   2445 assets  appear for more than 2300 days out of the total of 2356 trading days, and we focus on a random subset of $N=2048$ from this group.   On each day $t$ an asset $\rm i$ opens at price $p_{\rm i}^{\rm open}(t)$ and closes at price $p_{\rm i}^{\rm close}(t)$; we define the state of the system on day $t$ by the vector of daily returns $\{r_{\rm i}(t)\}$, with  $r_{\rm i}(t)=\ln[ p_{\rm i}^{\rm close}(t)/p_{\rm i}^{\rm open}(t)]$.  At left in Fig  \ref{stock_data} we see the spectrum of eigenvalues of the correlation matrix. In contrast to the neural data, the number of samples here is comparable to the dimensionality of the system, so we expect that the spectrum will be substantially affected by random sampling.   Indeed, if we project out the ten percent of modes with the largest variance (opposite to our RG procedure), the resulting spectrum is very close to the predictions of the Marchenko--Pastur distribution for  covariance matrices constructed from samples of uncorrelated variables \cite{JPB}.

If we apply our RG procedure to the raw data,  we see a non--monotonic trajectory of the fourth moments, first moving toward the Gaussian fixed point and then away (Fig \ref{stock_data}, right).    The turning point is roughly when we have integrated out all but the last ten percent of high variance modes, which is consistent with the lower $\sim 90\%$ of the eigenvalue spectrum being well described by the Marchenko--Pastur distribution.  Indeed, if we first exclude the top ten percent of high variance modes, the fourth moments flow very quickly to the Gaussian fixed point and the third moments flow to zero (not shown).  The ten percent of high variance modes clearly are not just noise, however, although it is not clear from the data whether their distribution is described by a fixed point of the RG.  We emphasize that the boundary between noise--like and non--noise modes is a property not of the system, but of the finite sample of data; it is possible that the RG analysis we propose here could be combined with denoising \cite{bun+al_15,bun+al_16} to give more insight.

\section{Not quite conclusions}

The idea that the RG might be useful in more complex systems is a widely held intuition.    In particular, there have been efforts to move from regular lattices to graphs \cite{on_graphs}, as well as  to construct a real space renormalization group for spin glasses \cite{castellana_11,angelini+al_13,angelini+biroli_15}.  What is new here, we think, is that, at least in perturbation theory, we can free ourselves completely from assumptions of locality, which seem so crucial to the usual notions of relevant and irrelevant operators.  Perhaps more importantly, connecting RG and PCA allows us to look at data in a new way, with interesting results in two very different complex systems.

PCA is a search for simplification.  The hope is that a system with variables that live in a high dimensional space can be captured by a projection of these variables in to a low dimensional space.  Although the RG involves (repeated) projections onto lower dimensional spaces, this dimensionality reduction is {\em not} the source of simplification. Indeed, when we study models,  the full renormalization group transformation involves expanding the system back to restore the original number of degrees of freedom; admittedly, this is difficult to do with real data.   RG is a search for simplification, not in the space of the system variables but in the space of models.  

An interesting connection is to the question of how well different terms in a model are determined by experimental data.  Starting with models for biological signaling networks \cite{brown+al_04}, Sethna and colleagues have argued that models for complex systems typically have a wide range of parameter sensitivities, so that some directions in parameter space have coordinates that are easily determined by data while other directions are almost never determined.  This pattern is quantified by the spectrum of eigenvalues in the Fisher information matrix (FIM), and in many cases this spectrum is nearly uniform on a logarithmic scale \cite{waterfall+al_06,gutenkunst+al_07}, a property termed ``sloppiness.''  Many of these models can be written so that parameter spaces are compact, and simplification then is achievable by moving along the ill--determined directions until reaching the edge of the space, leaving a model with one less parameter \cite{transtrum+al_11}.  Recent work has shown that conventional statistical physics models do not exhibit sloppiness if experiments involve measurements on the microscopic scale, but that this pattern develops when measurements are restricted to coarse--grained variables \cite{machta+al_13}.  The spreading of the FIM eigenvalues is controlled by the RG scaling of the different operators out of which the model is constructed, suggesting that the notions of simplification that are inherent to the RG are equivalent to a more data--driven simplification in which we keep only model components that are well determined be experiment.  It is possible that there are even more direct connections between the renormalization group and the learning of probabilistic models \cite{mehta+schwab_15}.

In the conventional implementations of the renormalization group, we put variables in order by their length scale, with small length scales at one end and long length scales at the other.  The intuition is that, when interactions are local, smaller scales are less important, or at least less interesting, and so we average over scales shorter than some distance $\ell$.  The RG then is the exploration of what happens as we change $\ell$.  In more complex systems, simplification requires us to find a  natural coordinate system in state space, and then put these coordinates in order of their likely importance, with fine--grained details at one end and crucial collective degrees of freedom at the other.  The spectrum of the covariance matrix gives us one possible answer to these questions, which we have explored here, but  surely there are other possibilities, even in the two examples discussed above.  The more  significant idea is that once we have identified an axis along which coarse graining seems to make sense, rather than looking for the right place to put the boundary between what we include and what we ignore, we should use the RG as inspiration to explore the evolution of our description as we move this boundary.

\begin{acknowledgments}
We thank D Amodei, MJ Berry II, and O Marre for making available the data of Ref \cite{tkacik+al_14} and M Marsili for the data of Ref \cite{Marsili2002}.  We are especially grateful to G Biroli, J--P Bouchaud, MP Brenner, CG Callan, A Cavagna, I Giardina, MO Magnasco, A Nicolis, SE Palmer, G Parisi, and DJ Schwab for helpful discussions and comments on the manuscript.
Work at CUNY was supported in part by the Swartz Foundation.  Work at Princeton was supported in part by grants from the National Science Foundation (PHY--1305525, PHY--1451171, and CCF--0939370) and the Simons Foundation.
\end{acknowledgments}

\end{document}